\documentclass[aps,prd,nofootinbib]{revtex4}
\usepackage{amsmath}
\usepackage{graphicx}
\usepackage{dcolumn}
\usepackage{bm}
\usepackage{amssymb}
\usepackage{latexsym}

\begin{document}

\title{Bouncing Universe with Quintom Matter }

\author{Yi-Fu Cai$^a$\footnote{caiyf@mail.ihep.ac.cn}, Taotao Qiu$^a$\footnote{qiutt@mail.ihep.ac.cn},
Yun-Song Piao$^b$, Mingzhe Li$^c$, Xinmin Zhang$^a$}

\vspace{1.cm}
\address{$^a$Institute of High Energy Physics, Chinese Academy
of Sciences, P.O. Box 918-4, Beijing 100049, P. R. China
\vspace{.1in} \\
$^b$College of Physical Sciences, Graduate School of Chinese
Academy of Sciences, YuQuan Road 19A, Beijing 100049, China
\vspace{.1in}\\
$^c$Fakult\"{a}t f\"{u}r Physik, Universit\"{a}t Bielefeld,
D-33615 Bielefeld, Germany}

\begin{abstract}

The bouncing universe provides a possible solution to the Big Bang
singularity problem. In this paper we study the bouncing solution
in the universe dominated by the Quintom matter with an equation
of state (EoS) crossing the cosmological constant boundary. We
will show explicitly the analytical and numerical bouncing
solutions in three types of models for the Quintom matter with an
phenomenological EoS, the two scalar fields and a scalar field
with a modified Born-Infeld action.

\end{abstract}

\maketitle


\section{Introduction}

A bouncing universe with an initial contraction to a non-vanishing
minimal radius, then subsequent an expanding phase provides a
possible solution to the singularity problem of the standard Big
Bang cosmology. For a successful bounce, it can be shown that
within the framework of the standard 4-dimensional
Friedmann-Robertson-Walker (FRW) cosmology with Einstein gravity
the null energy condition (NEC) is violated for a period of time
around the bouncing point. Moreover, for the universe entering
into the hot Big Bang era after the bouncing, the EoS of the
matter content $w$ in the universe must transit from $w<-1$ to
$w>-1$.

The Quintom model \cite{Quintom1}, proposed to understand the
behavior of dark energy with an EoS of $w>-1$ in the past and
$w<-1$ at present, has been supported by the observational
data\cite{Zhao:2006qg}. Quintom is a dynamical model of dark
energy. It differs from the cosmological constant, Quintessence,
Phantom, K-essence and so on in the determination of the
cosmological evolution. A salient feature of the Quintom model is
that its EoS can smoothly cross over $w=-1$. In the recent years
there has been a lot of proposals for the Quintom-like models in
the literature. In this paper we study the bouncing solution in
the universe dominated by the Quintom matter and working with
three specific models we will show explicitly the analytical and
numerical solutions of the bounce.

We will start with a detailed examination on the necessary
conditions required for a successful bounce. During the
contracting phase, the scale factor $a(t)$ is decreasing, i.e.,
$\dot a(t)<0$, and in the expanding phase we have $\dot a(t)>0$.
At the bouncing point, $\dot a(t)=0$, and around this point $\ddot
a(t)>0$ for a period of time. Equivalently in the bouncing
cosmology the hubble parameter $H$ runs across zero from $H<0$ to
$H>0$ and $H=0$ at the bouncing point. A successful bounce
requires around this point,
\begin{eqnarray}\label{bccong}
\dot H=-4\pi G\rho (1+w)>0~.
\end{eqnarray}
From (\ref{bccong}) one can see that $w<-1$ in a neighborhood of
the bouncing point.

After the bounce the universe needs to enter into the hot Big Bang
era, otherwise the universe filled with the matter with an EoS
$w<-1$ will reach the big rip singularity as what happens to the
Phantom dark energy\cite{Caldwell:2003vq}. This requires the EoS of
the matter to transit from $w<-1$ to $w>-1$.

In this paper, we study the bouncing solutions in the Quintom
models. The paper is organized as follows. In section II, we
present the analytical and numerical solutions for different types
of models of the Quintom matter. Specifically we consider three
models: i)a phenomenological Quintom fluid with a parameterized
EoS crossing the cosmological constant boundary; ii)the two-field
models of Quintom matter with one being the quintessence-like
scalar and another the phantom-like scalar; iii) a single scalar
with a Born-Infeld type action. III is the summary of the paper.

\section{Bouncing solution in the presence of Quintom matter}

\subsection{A phenomenological Quintom model}

We start with a study on the possibility of obtaining the bouncing
solution in a phenomenological Quintom matter described by the
following EoS:
\begin{equation}\label{paratachyon}
w(t)=-r-\frac{s}{t^2}~.
\end{equation}
In (\ref{paratachyon}) $r$ and $s$ are parameters and we require
that $r<1$ and $s>0$. One can see from (\ref{paratachyon}) that $w$
runs from negative infinity at $t=0$ to the cosmological constant
boundary at $t=\sqrt{\frac{s}{1-r}}$ and then crosses this boundary.

\begin{widetext}
\begin{figure}[htbp]
\includegraphics[scale=0.7]{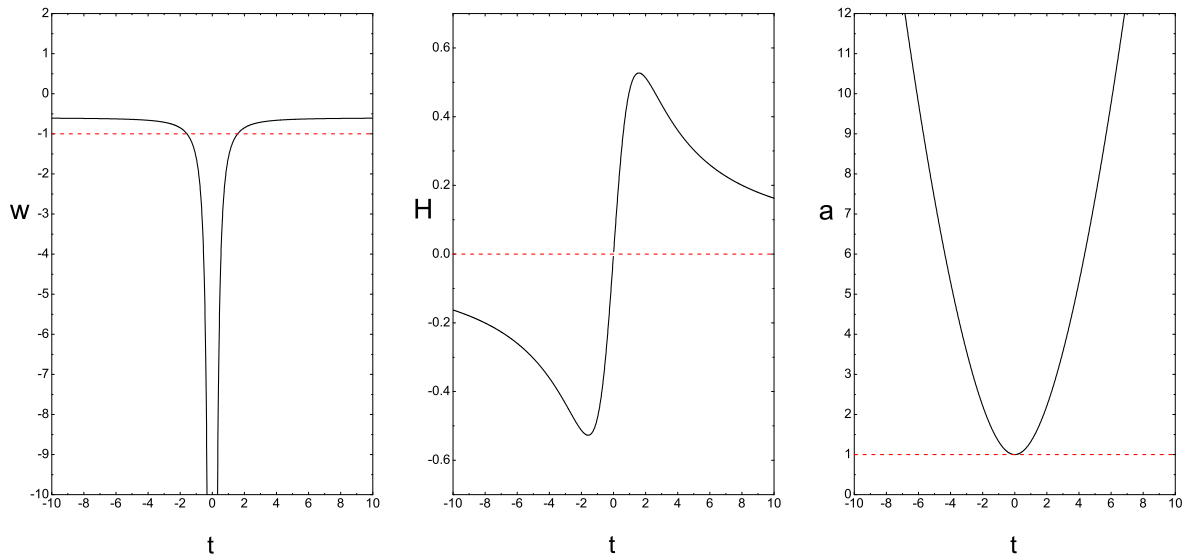}
\caption{Plot of the evolution of the EoS $w$, the hubble
parameter $H$ and the scale factor $a$ as a function of the cosmic
time $t$. Here in the numerical calculation we have taken $r=0.6$
and $s=1$.} \label{fig:1parahw}
\end{figure}
\end{widetext}

Assuming that the universe is dominated by the matter with the EoS
given by (\ref{paratachyon}), we solve the Friedmann equation and
obtain the corresponding evolution of hubble parameter $H(t)$ and
scale factor $a(t)$ as follows,
\begin{eqnarray}
H(t)&=&\frac{2}{3}\frac{t}{(1-r) t^2+s}~,\\
a(t)&=&(t^2+\frac{s}{1-r})^{\frac{1}{3 (1-r)}}~.
\end{eqnarray}
Here we choose $t=0$ as the bouncing point and normalize $a=1$ at
this point. One can see that our solution provides a picture of
the universe evolution with contracting for $t<0$, and then
bouncing at $t=0$ to the expanding phase for $t>0$. In Fig.
\ref{fig:1parahw} we plot the evolution of the EoS, the hubble
parameter and the scale factor.

One can see from Fig. \ref{fig:1parahw} that a non-singular
bouncing happens at $t=0$ with the hubble parameter $H$ running
across zero and a minimal non-vanishing scale factor $a$. At the
bouncing point $w$ approaches negative infinity.

\subsection{Two-field Quintom model}

Having presented the bouncing solution with the phenomenological
Quintom matter, we now study the bounce in the scalar field models of
Quintom
matter. However it is not easy to build a Quintom model
theoretically. The No-Go theorem proven in Ref. \cite{xiacai07}
(also see Ref.
\cite{Quintom1,zhaogbper,Caldwell05,Vikman05,wHu05,Kunz06})
forbids the traditional scalar field model with a lagrangian of
general form ${\cal L}={\cal L}(\phi,
\nabla_\mu\phi\nabla^\mu\phi)$ to have its EoS cross over the
cosmological constant boundary. Therefore, to realize a viable
Quintom field model in the framework of Einstein's gravity theory,
it needs to introduce extra degree of freedom to the conventional
theory with a single scalar field. The simplest
Quintom model involves two scalars with one being the Quintessence-like
and another the Phantom-like
\cite{Quintom1,Guozk}. This model has been studied
in detail later on in the literature. In the recent years there have been
a lot of activities in the theoretical study on
Quintom-like models such as a single scalar with high-derivative
\cite{lfz,arefeva}, vector field\cite{Wei}, extended theory
of gravity\cite{extendgravity} and so on, see e.g.
\cite{Quintomsum}.

In this section we consider a two-field Quintom model with the
action given by
\begin{eqnarray}
S=\int d^4x \sqrt{-g}
\left[\frac{1}{2}\partial_{\mu}\phi_1\partial^{\mu}\phi_1-\frac{1}{2}\partial_{\mu}\phi_2\partial_{\mu}\phi_2-V(\phi_1,\phi_2)\right]~,
\end{eqnarray}
where the metric is in form of $(+,-,-,-)$. Here the field
$\phi_1$ has a canonical kinetic term, but $\phi_2$ is a ghost field. In
the framework of FRW cosmology, we can easily obtain the energy
density and the pressure of this model,
\begin{eqnarray}
\rho=\frac{1}{2}{\dot\phi_1}^2-\frac{1}{2}{\dot\phi_2}^2+V~,~~~~
p=\frac{1}{2}{\dot\phi_1}^2-\frac{1}{2}{\dot\phi_2}^2-V~,
\end{eqnarray}
and the Einstein equations are given by
\begin{eqnarray}
\label{Einstein1}
H^2=\frac{8\pi G}{3}(\frac{1}{2}{\dot\phi_1}^2-\frac{1}{2}{\dot\phi_2}^2+V)~,\\
\label{Einstein2}
\ddot\phi_1+3H\dot\phi_1+\frac{dV}{d\phi_1}=0~,\\
\label{Einstein3} \ddot\phi_2+3H\dot\phi_2-\frac{dV}{d\phi_2}=0~.
\end{eqnarray}

From Eq. (\ref{bccong}), we can see that a bouncing solution
requires ${\dot\phi_2}^2={\dot\phi_1}^2+2V$ when $H$ crosses zero;
and the Quintom behavior requires ${\dot\phi_2}^2={\dot\phi_1}^2$
when $w$ crosses $-1$. These constraints can be easily satisfied
in the parameter space of this model.

\begin{figure}[htbp]
\includegraphics[scale=0.6]{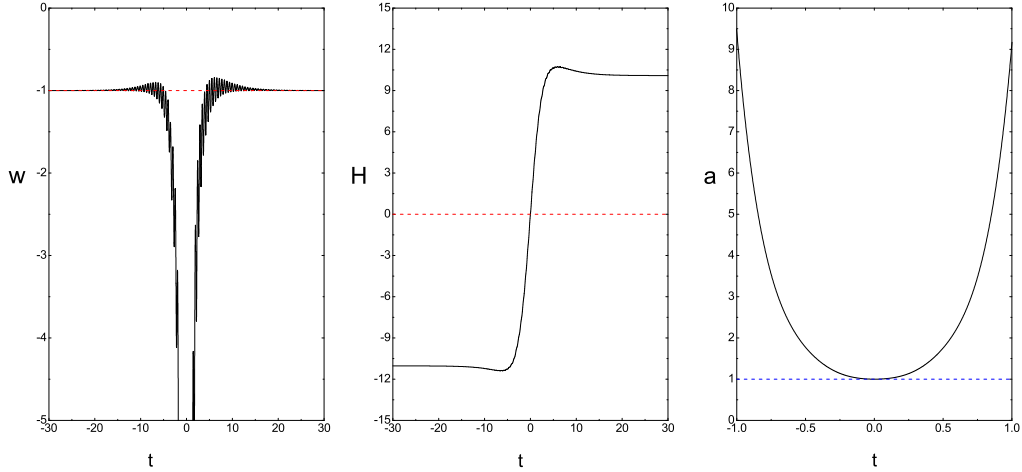}
\caption{The plots of the evolutions of the EoS $w$, hubble
parameter $H$ and the scale factor $a$. In the numerical
calculation we choose $V(\phi_1,\phi_2)=V_1
e^{-\lambda_1\frac{\phi_1^2}{M^2}}+V_2
e^{-\lambda_2\frac{\phi_2^2}{M^2}}$ with parameters:
$V_1=15,~V_2=1,~\lambda_1=-1.0,~\lambda_2=1.0$, and for the
initial conditions
$\phi_1=0.5,~\dot\phi_1=0.1,~\phi_2=0.3,~\dot\phi_2=4$.}
\label{fig:double}
\end{figure}

\begin{figure}[htbp]
\includegraphics[scale=0.6]{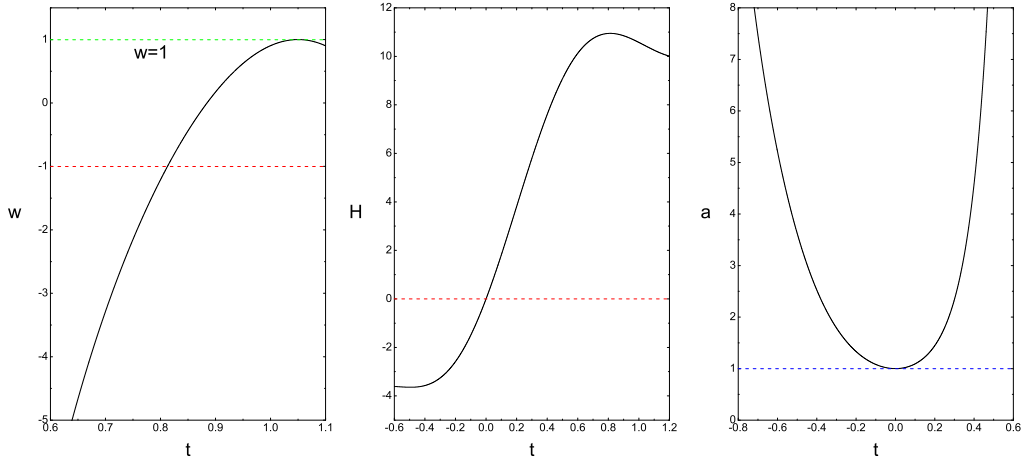}
\caption{The same plots as Fig. \ref{fig:double} with different
potential and model parameters
$V(\phi)=\frac{1}{2}m^2{\phi_1}^2+V_0{\phi_2}^{-2},~m=2,~V_0=0.4$,
and for the initial conditions
$\phi_1=2,~\dot\phi_1=3,~\phi_2=1,~\dot\phi_2=2$.}
\label{fig:doublew1}
\end{figure}

In Fig. \ref{fig:double} and Fig. \ref{fig:doublew1}, we show the
bouncing solution for two different type of potentials. In Fig.
\ref{fig:double} we take $V(\phi_1,\phi_2)=V_1
e^{-\lambda_1\frac{\phi_1^2}{M^2}}+V_2
e^{-\lambda_2\frac{\phi_2^2}{M^2}}$. In the numerical calculation
we normalize the dimensional parameters such as $V_1$, $V_2$,
$\phi_1$ and $\phi_2$ by a mass scale $M$ which we take
specifically to be $10^{-2}M_{pl}$. And the hubble parameter is
normalized with $\frac{M^2}{M_{pl}}$. One can see from this figure
the non-singular behavior of the Hubble parameter and the scale
factor for a bounce. The EoS $w$ crosses over the $w=-1$ and
approaches to negative infinity at the bouncing point. And due to
the oscillatory behavior of the field $\phi_1$ in the evolution,
the EoS $w$ is also oscillating around bouncing point.

In Fig. \ref{fig:doublew1}, we take
$V(\phi)=\frac{1}{2}m^2{\phi_1}^2+V_0{\phi_2}^{-2}$.
 This model also provides a
bouncing solution, however the detailed evolution of the universe
differs from the one shown in Fig. \ref{fig:double}. Fig.
\ref{fig:doublew1} shows that the EoS of the Quintom matter will
approach $w=1$ asymptotically.

\subsection{A single scalar with high-derivative terms}

In this section we consider a class of Quintom models described by
an effective lagrangian with higher derivative operators. Starting
with a canonical scalar field with the lagrangian ${\cal L}=
\frac{1}{2}
\partial_\mu \phi \partial^\mu \phi - V(\phi)$. This type of
models has been considered as a candidate for dark energy, however
as shown by the No-Go theorem it does not give $w$ crossing -1. As
an effective theory we know that the lagrangian should include
more operators, especially if these operators involve the term
$\Box \phi$, as pointed in Ref. \cite{lfz} it will give rise to an
EoS across $w=-1$.  A connection of this type of Quintom theory to
the string theory has been considered in Ref. \cite{cyftachyon}
and \cite{arefeva}. In this paper we take the string-inspired
model in \cite{cyftachyon} for the detailed study on the bouncing
solution, where the action is given by
\begin{eqnarray}\label{tachyonaction}
S=\int d^4x\sqrt{-g}\left[-V(\phi)\sqrt{1-{\alpha}^\prime
\nabla_{\mu}\phi\nabla^{\mu}\phi+{\beta}^\prime\phi\Box\phi}\right]~.
\end{eqnarray}
This is a generalized version of ``Born-Infeld"
action\cite{Gerasimov:2000zp, Kutasov:2000qp} with the
introduction of the $\beta^\prime$ term. To the lowest order, the
Box-operator term $\phi\Box\phi$ is equivalent to the term
$\nabla_{\mu}\phi\nabla^{\mu}\phi$ when the tachyon is on the top
of its potential. However when the tachyon rolls down from the top
of the potential, these two terms exhibit different dynamical
behavior. The two parameters $\alpha'$ and $\beta'$ in
(\ref{tachyonaction}) could be arbitrary in the case of the
background flux being turned on \cite{Mukhopadhyay:2002en}. One
interesting feature of this model is that it provides the
possibility of its EoS $w$ running across the cosmological
constant boundary. In the analytical and numerical studies below
to make two parameters ($\alpha^\prime$, $\beta^\prime$)
dimensionless, it is convenient to redefine $ \alpha
=\alpha^\prime M^4$ and $\beta = \beta^\prime M^4$ where $M$ is an
energy scale of the effective theory of tachyon.

From (\ref{tachyonaction}) we obtain the equation of motion for the
scalar field $\phi$:
\begin{eqnarray}\label{eqom}
\frac{\beta}{2}\Box(\frac{V\phi}{f})+
\alpha\nabla_{\mu}(\frac{V\nabla^{\mu}\phi}{f})+M^4V_{\phi}f+\frac{\beta
V}{2f}\Box\phi=0~,
\end{eqnarray}
where $f=\sqrt{1-{\alpha}^\prime
\nabla_{\mu}\phi\nabla^{\mu}\phi+{\beta}^\prime\phi\Box\phi}$ and
$V_\phi=dV /d\phi$. Correspondingly, the stress energy tensor of
the model is given by
\begin{eqnarray}\label{stress}
T_{\mu\nu}=g_{\mu\nu}[Vf-\frac{\beta}{2M^4}\nabla_{\rho}(\frac{\phi
V}{f}\nabla^{\rho}\phi)]+
\frac{\alpha}{M^4}\frac{V}{f}\nabla_\mu\phi\nabla_\nu\phi+\frac{\beta}{2M^4}\nabla_\mu(\frac{\phi
V}{f})\nabla_\nu\phi+\frac{\beta}{2M^4}\nabla_\nu(\frac{\phi
V}{f})\nabla_\mu\phi~.
\end{eqnarray}
Technically, it is very useful to define a parameter
$\psi\equiv\frac{\partial{\cal
L}}{\partial\Box\phi}=-\frac{\beta\phi V}{2M^4f}$ to solve
(\ref{eqom}) and (\ref{stress}). In the framework of a flat FRW
universe filled with a homogenous scalar field $\phi$, we have the
equations of motion in form of
\begin{eqnarray}
\ddot\phi+3H\dot\phi&=&\frac{\beta\phi}{4M^4\psi^2}V^2-\frac{M^4}{\beta\phi}+\frac{\alpha}{\beta\phi}\dot\phi^2~,\\
\ddot\psi+3H\dot\psi&=&(2\alpha+\beta)(\frac{M^4\psi}{\beta^2\phi^2}-\frac{V^2}{4M^4\psi})-\frac{\beta\phi}{2M^4\psi}VV_{\phi}
-(2\alpha-\beta)\frac{\alpha\psi}{\beta^2\phi^2}\dot\phi^2-\frac{2\alpha}{\beta\phi}\dot\psi\dot\phi~.
\end{eqnarray}
Moreover, the energy density and the pressure of this field can be
written as
\begin{eqnarray}
\label{rho}\rho=-\frac{\alpha\psi}{\beta\phi}\dot\phi^2-\dot\psi\dot\phi-\frac{\beta\phi}{4M^4\psi}V^2-\frac{M^4\psi}{\beta\phi}~,\\
\label{press}
p=-\frac{\alpha\psi}{\beta\phi}\dot\phi^2-\dot\psi\dot\phi+\frac{\beta\phi}{4M^4\psi}V^2+\frac{M^4\psi}{\beta\phi}~.
\end{eqnarray}

From Eq.(\ref{bccong}), one can see that a successful bounce
requires:
\begin{eqnarray}\label{bctachyon}
\frac{\beta\phi}{4M^4\psi}V^2+\frac{M^4\psi}{\beta\phi}=
-\frac{\alpha\psi}{\beta\phi}\dot\phi^2-\dot\psi\dot\phi<0~.
\end{eqnarray}
We will show below that (\ref{bctachyon}) can be satisfied easily
for our model. In the numerical study on the bouncing solution, we
constrain the parameters $\alpha$ and $\beta$ so that the model in
(\ref{tachyonaction}) when expanding the derivative terms in the
square root to the lowest order gives rise to a canonical kinetic
term for the scalar field $\phi$ \cite{cyftachyon}, i.e.,
$\alpha+\beta>0$.

\begin{figure}[htbp]
\includegraphics[scale=0.6]{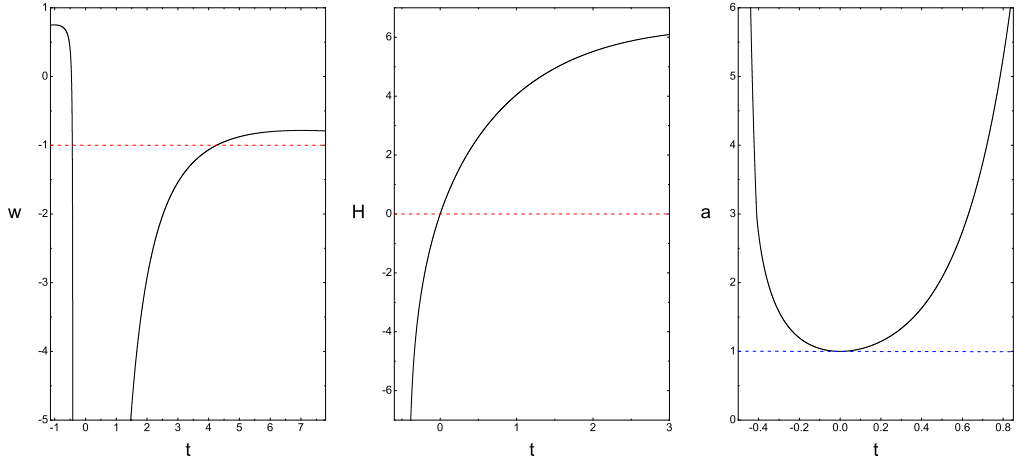}
\caption{The plots of the evolution of the EoS $w$, the hubble
parameter $H$ and the scale factor $a$. Here in the numerical
calculation we take the potential $V(\phi)=V_0e^{-\lambda\phi^2}$,
$\alpha=-0.2$, $\beta=2$, $\lambda=2$, $V_0=5$, and the initial
values are $\phi=1$, $\dot\phi=3$, $H=-1$, and $\psi=-80$.}
\label{fig:bouncinginf}
\end{figure}

\begin{figure}[htbp]
\includegraphics[scale=0.6]{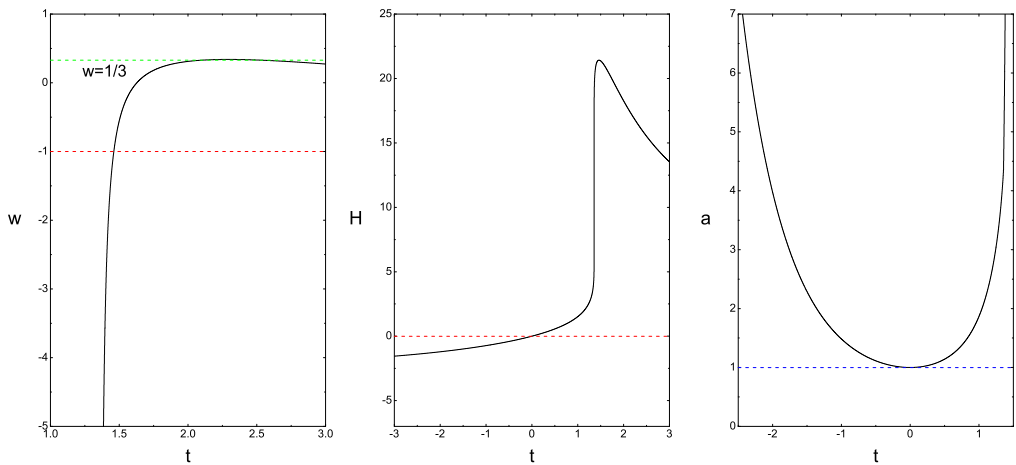}
\caption{The plots of the evolutions of the EoS $w$, hubble
parameter $H$ and the scale factor $a$. In the numerical
calculation we choose the potential as $V(\phi)=\frac{V_0}{\phi}$,
$\alpha=-0.2,~\beta=2,~V_0=0.7$, and for the initial conditions
$\phi=10,~\dot\phi=-3,~H=-1,~\psi=-40$.} \label{fig:bouncingrad}
\end{figure}

In Fig. \ref{fig:bouncinginf} and Fig. \ref{fig:bouncingrad} we
show the bounce solution for different potentials. In Fig.
\ref{fig:bouncinginf}, we take $V(\phi)=V_0e^{-\lambda\phi^2/M^2}$
with $\lambda$ being a dimensionless parameter.  One can see from
this figure the scale factor initially decreases, then passes
through its minimum and increases. Moreover, away from the
bouncing point in the expanding phase, the EoS of the scalar field
crosses $w=-1$ and approaches $w=-0.6$, which gives rise to a
possible inflationary phase after the bouncing. In the numerical
calculation we take the energy scale $M$ to be $10^{-2}M_{pl}$,
and the hubble parameter is normalized with $\frac{M^2}{M_{pl}}$.

In Fig. \ref{fig:bouncingrad}, we consider the model with
potential $V(\phi)=V_0/\phi$ and then show another example of the
bouncing solution\footnote{A Born-Infeld lagrangian with this
potential provides a scaling solution, see Ref.
\cite{Tsujikawascaling}.}. Here the energy scale $M$ is chosen  to
be $10^{-2}M_{pl}$ as well. One can see from this figure the clear
picture of the bouncing, however the detailed evolution of the
universe differs from the one shown in Fig. \ref{fig:bouncinginf}.
After entering the expanding phase, the EoS $w$ crosses the
cosmological constant boundary and approaches $w=\frac{1}{3}$,
which is equivalent to the EoS of the radiation.

\section{Conclusion and discussions}

In this paper we have studied the possibility of obtaining a
non-singular bounce in the presence of the Quintom matter. In the
literature there have been a lot of efforts in constructing the
bouncing universe, for instance, the Pre Big Bang
scenario\cite{GV}, and the Ekpyrotic scenario\cite{Steinhardt01}.
In Refs. \cite{BM, CCM,TBF} and \cite{BMS, BBMS} the authors have
considered models with the modifications of gravity with the high
order terms. In general these models modify the 4-dimensional
Einstein gravity. However, the models we consider for bounce
universe in this paper are restricted to be within the standard
4-dimensional FRW framework.

Recently two papers \cite{bounceghost, bounceghost2} have studied
the possibilities of having a bounce universe with the ghost
condensate. In the original formulation the ghost
condensate\cite{Arkani-Hamed:2003uy} will not be able to give EoS
crossing $w=-1$. The authors of these papers\cite{bounceghost,
bounceghost2} have considered a generalized model of ghost
condensate\cite{Tsujikawa05} and shown the bouncing solutions. In
this paper we have studied the general issue of obtaining a
bouncing universe with the Quintom matter. Our results show that a
universe in the presence of the Quintom matter will avoid the
problem of the Big Bang singularity. Explicitly for the analytical
and numerical studies we have considered three models: the
phenomenological model, the two-field model and the
string-inspired Quintom model. The latter one is a generalization
of the idea in Ref.\cite{lfz} by introducing higher derivative
terms to realize the EoS crossing $w=-1$. In this regard, this
model for the bounce solution has the similarity with a recent
paper \cite{eva07} where the authors presented a bouncing solution
with non-local SFT\cite{arefeva}.

\acknowledgments

We thank Jie Liu, Jian-Xin Lu, Anupam Mazumdar, Shinji Tsujikawa,
Jun-Qing Xia, and Gong-Bo Zhao, for discussions. This work is
supported in part by National Natural Science Foundation of China
under Grant Nos. 90303004, 10533010 and 10675136 and by the
Chinese Academy of Science under Grant No. KJCX3-SYW-N2.

\end{document}